\documentclass[12pt,eqsecnum]{article}
\usepackage[dvips]{graphicx}
\usepackage{amssymb}
\usepackage{amsmath}
\usepackage{ulem}
\usepackage[dvipsnames]{xcolor}
\usepackage{soul}
\usepackage{cite}
\makeatletter

\@addtoreset{equation}{section}
\makeatletter

\newcommand{\qed}{\hbox{\rule[-2pt]{6pt}{6pt}}}
\newcommand{\D}{{\rm d}}

\newtheorem{Prop}{Proposition}

\newcommand{\dalm}{\kern1pt\vbox{\hrule height 0.9pt\hbox{\vrule width
0.9pt\hskip 2.5pt\vbox{\vskip 5.5pt}\hskip 3pt\vrule width 0.3pt}\hrule height
0.3pt}\kern1pt}

\begin{document}

\begin{titlepage}
\vfill
\begin{flushright}
\today
\end{flushright}

\vfill
%\vskip 1.0cm
\begin{center}
\baselineskip=16pt
{\Large\bf 
Fake Schwarzschild and Kerr black holes
}
\vskip 0.5cm
{\large {\sl }}
\vskip 10.mm
{\bf Hideki Maeda} \\

\vskip 1cm
{
Department of Electronics and Information Engineering, Hokkai-Gakuen University, Sapporo 062-8605, Japan.\\
\texttt{h-maeda@hgu.jp}

}
\vspace{6pt}
%\today
\end{center}
\vskip 0.2in
\par
\begin{center}
{\bf Abstract}
\end{center}
\begin{quote}
We present exact solutions describing a {\it fake Schwarzschild} black hole that cannot be distinguished from the Schwarzschild black hole by observations.
They are constructed by attaching a spherically symmetric dynamical interior solution with a matter field to the Schwarzschild exterior solution at the event horizon without a lightlike thin shell.
The dynamical region inside a Killing horizon of a static spherically symmetric perfect-fluid solution obeying an equation of state $p=\chi\rho$ for $\chi\in[-1/3,0)$ can be the interior of a fake Schwarzschild black hole.
The matter field inside such a black hole is an anisotropic fluid that violates at least the weak energy condition and can be interpreted as a spacelike (tachyonic) perfect fluid.
While the author constructed the first model of fake Schwarzschild black holes using Semiz's solution for $\chi=-1/5$, we present another one using Whittaker's solution for $\chi=-1/3$ in this paper.
We also present a model of {\it fake Kerr} black holes whose interior is filled with a different matter field violating only the dominant energy condition.
Because it contradicts the conservation theorem, this configuration of black holes is, in fact, precluded by the dominant energy condition.
\vfill
% \hrule width 5.cm
\vskip 2.mm
\end{quote}
\end{titlepage}

%<<<<<<<<<<<<< PACS NUMBER >>>>>>>>>>>>>>>%
%\pacs{
%04.20.]q Classical general relativity
%04.20.Jb Exact solutions
%04.40.]b Self-gravitating systems; continuous media and classical fields in curved spacetime
%04.70.Bw Classical black holes
%} 

% CECS-PHY-13/09

%\maketitle
%\section{}
%\subsection{}

\tableofcontents

\newpage

%======================================%
%<<<<<<<<<<<< SECTION I >>>>>>>>>>>>>>%
%======================================%
\section{Introduction}

By Birkhoff's theorem, the Schwarzschild solution is the only spherically symmetric vacuum solution to the Einstein equations without a cosmological constant.
It describes an asymptotically flat black hole with a single non-degenerate Killing horizon corresponding to an event horizon when the Arnowitt-Deser-Misner (ADM) mass is positive.
Undoubtedly, the Schwarzschild solution is the simplest model of black holes in our universe and is the basis for the study of black hole physics.
Nevertheless, it is not a trivial question whether the interior of a realistic black hole is described by the Schwarzschild solution.
It is often claimed that the shadow of a black hole has been observed by gravitational lensing~\cite{EventHorizonTelescope:2019dse}, however, the radius of the shadow is larger than the radius of the event horizon of the Schwarzschild black hole.
Therefore, one cannot immediately conclude that the Schwarzschild spacetime can approximate the interior of the shadow, or even that an event horizon exists there.

For example, Mazur and Mottola proposed a final state of gravitational collapse that is not a black hole, called a {\it gravastar}~\cite{Mazur:2004fk,Mazur:2001fv}.
A gravastar spacetime consists of three regions, where the inner de~Sitter region I and the outer Schwarzschild region III are connected through an intermediate region II filled with a stiff fluid.
As a result, a gravastar spacetime does not admit an event horizon even though it is very similar to the Schwarzschild black hole.
However, in exchange for that, massive thin shells exist at the boundaries between regions I and II, and between regions II and III, where the first derivative of the metric is discontinuous.
A massive thin shell is a curvature singularity, although it can be treated as a solution to the Einstein equations as a distribution~\cite{Poisson:2009pwt}.

The author and Mart\'{\i}nez recently showed that the Schwarzschild vacuum interior solution can be attached on the Killing horizon regularly without a lightlike massive thin shell to a static spherically symmetric exterior solution involving a matter field~\cite{Maeda:2024lbq}.
If linear relations $p_{\rm r}\simeq \chi_{\rm r}\rho$ and $p_{\rm t}\simeq \chi_{\rm t}\rho$ are satisfied near a Killing horizon for the energy density $\rho$, radial pressure $p_{\rm r}$, and tangential pressure $p_{\rm t}$ of the matter field, such a regular attachment is possible for $\chi_{\rm r}\in[-1/3,0)$.
This result shows that, if a static spherically symmetric solution with a perfect fluid obeying $p=\chi\rho$ for $\chi\in[-1/3,0)$ admits a Killing horizon, where $p(=p_{\rm r}=p_{\rm t})$ is the pressure of the fluid, it can be attached to the Schwarzschild interior solution there.
The resultant spacetime is an exact solution to the Einstein equations, and the metric is not smooth on the horizon, however, there is no lightlike massive thin shell.

In the above construction, the interior and exterior spacetimes can be interchanged.
That is, the Schwarzschild exterior solution can be attached regularly on the Killing horizon to the dynamical region inside a Killing horizon of a static spherically symmetric perfect-fluid solution for $\chi\in[-1/3,0)$.
We will refer to such a configuration of black holes as a {\it fake Schwarzschild black hole}, which cannot be distinguished from the Schwarzschild black hole by external observers in principle.
Although the result in Ref.~\cite{Maeda:2024lbq} shows that fake Schwarzschild black holes are possible using perfect-fluid solutions for $\chi\in[-1/3,0)$, we need to construct such solutions to clarify the spacetime structure and the properties of the matter field inside the horizon.
In particular, it is an important question whether the matter field that allows such a fake black hole can satisfy all the standard energy conditions, and if not, to what extent they can be satisfied.
This paper gives a partial answer to this question.
In Ref.~\cite{Maeda:2022lsm}, the author constructed the first example of exact solutions describing a fake Schwarzschild black hole using Semiz's solution for $\chi=-1/5$~\cite{Semiz:2020lxj}\footnote{In Ref.~\cite{Maeda:2022lsm}, the author also constructed exact solutions describing a fake Tangherlini black hole in five dimensions using the five-dimensional counterpart of Semiz's solution for $\chi=-1/3$.}.
In the present paper, we will construct another example using Whittaker's solution for $\chi=-1/3$~\cite{Whittaker}.
For $\chi\in[-1/3,0)$, the matter field inside a fake Schwarzschild black hole in this construction is not a perfect fluid but an anisotropic fluid.
We will show that it can satisfy the null and strong energy conditions but violates the weak energy condition (and hence the dominant energy condition, too).
Nonetheless, we will show that such a configuration of black holes is possible under the WEC and SEC by constructing a model of {\it fake Kerr} black holes described by the G\"urses-G\"ursey metric~\cite{Gurses:1975vu} with a different matter field that is of the Hawking-Ellis type I and obeys $p_{\rm r}=-\rho$.
This model is a rotating generalization of a fake Schwarzschild black hole studied by Ovalle in Ref.~\cite{Ovalle:2024wtv}.

The organization of the present paper is as follows.
In Sec.~\ref{sec:preliminaries}, we review several results for static spherically symmetric perfect-fluid solutions and study the Whittaker solution for the subsequent section.
In Sec.~\ref{sec:main}, we present exact solutions describing a fake Schwarzschild black hole using the Semiz and Whittaker solutions as interior solutions.
In Sec.~\ref{sec:main2}, we present and study three different models for fake Kerr black holes.
We summarize our results and give concluding remarks in the final section.
In Appendix~\ref{app:EC-tachyon}, we derive equivalent representations of the standard energy conditions for a tachyonic perfect fluid in arbitrary $n(\ge 3)$ dimensions without assuming any spacetime symmetry.
In Appendix~\ref{app:topo}, we perform the same analysis presented in Sec.~\ref{sec:extension} to the topological generalization of the Whittaker solution with an arbitrary two-dimensional maximally symmetric base manifold in the presence of $\Lambda$.
Our conventions for curvature tensors are ${[\nabla _\rho ,\nabla_\sigma]V^\mu ={{R}^\mu }_{\nu\rho\sigma}V^\nu}$ and ${{R}_{\mu \nu }={{R}^\rho }_{\mu \rho \nu}}$, where Greek indices run over all spacetime indices.
The signature of the Minkowski spacetime is $(-,+,+,+)$, and we adopt the units such that $c=1$ and $\kappa:=8\pi G$, where $G$ is the gravitational constant.
The sets of natural numbers are denoted by $\mathbb{N}(=1,2,3,\cdots)$ and $\mathbb{N}_0(=0,1,2,\cdots)$.
Throughout this paper, a prime denotes differentiation with respect to the argument of the function.

%======================================%
%<<<<<<<<<<<< SECTION I >>>>>>>>>>>>>>%
%======================================%
\section{Preliminaries}
\label{sec:preliminaries}
In this section, we focus on static spherically symmetric solutions in general relativity with a perfect fluid.
The Einstein equations are written as
\begin{align}
\label{EFE}
\begin{aligned}
&G_{\mu\nu}=\kappa T_{\mu\nu},\\
&T_{\mu\nu}=\rho u_\mu u_\nu+p(g_{\mu\nu}+u_\mu u_\nu),
\end{aligned}
\end{align}
where $\rho$ and $p$ are the energy density and pressure of the fluid, respectively, and the four-velocity of the fluid element $u^\mu$ is normalized such that $u_\mu u^\mu=-1$.
We assume a linear equation of state $p=\chi\rho$, where $\chi$ is a constant.
The results in this section will be used in Sec.~\ref{sec:main}.

\subsection{Static spherically symmetric solutions with Killing horizons}

For static spherically symmetric spacetimes with a perfect fluid, we can use the comoving quasi-global coordinates $(t,x,\theta,\phi)$ without loss of generality such that
\begin{align}
\label{Semiz-I-twopara2}
\begin{aligned}
&\D s^2=-H(x)\D {t}^2+H(x)^{-1}\D x^2+r(x)^2\D\Omega^2,\\
&u^\mu\partial_\mu=H^{-1/2}\partial_t,
\end{aligned} 
\end{align} 
where $H>0$ and $\D\Omega^2:=\D\theta^2+\sin^2\theta\D\phi^2$.
The energy-momentum tensor of a perfect fluid (\ref{EFE}) is given in the coordinates (\ref{Semiz-I-twopara2}) as
\begin{align}
\label{T-diag}
T^\mu_{~\nu}=\mbox{diag}(-\rho,p,p,p).
\end{align} 
In the dust case ($\chi=0$), non-vacuum static spherically symmetric solutions do not exist. (See Appendix~B in Ref.~\cite{Maeda:2024lbq}.)
For $\chi=-1$, the energy-momentum conservation equations $\nabla_\nu T^{\mu\nu}=0$ show that the perfect fluid is equivalent to a cosmological constant $\Lambda(=\rho=-p)$.
Then, by Birkhoff's theorem with $\Lambda$, the general solution consists of the Schwarzschild-(anti-)de~Sitter solution for $\Lambda\ne 0$ and the Nariai solution for $\Lambda>0$ .

In the coordinate system (\ref{Semiz-I-twopara2}), $x=x_{\rm h}$ defined by $H(x_{\rm h})=0$ is a coordinate singularity.
If regular, it is a Killing horizon and the spacetime can be extended beyond there.
To avoid the coordinate singularity, we introduce a null coordinate $v:=t\pm \int H(x)^{-1}\D x$ and then the metric becomes
\begin{equation}
\D s^2=-H(x)\D v^2\pm 2\D v\D x+r(x)^2\D\Omega^2.\label{metric-Buchdahl-v}
\end{equation}
A regular null hypersurface $x=x_{\rm h}$ with the plus (minus) sign corresponds to a future (past) Killing horizon.

In general relativity, a $C^{1,1}$ metric, often denoted by $C^{2-}$ in physics, is sufficient for regularity as it avoids curvature singularities as well as the divergence of matter fields.
(See Sec.~2.3 in Ref.~\cite{Maeda:2024lbq}.)
A $C^{1,1}$ metric is continuously differentiable ($C^1$) and has the first derivative that is locally Lipschitz continuous, so that $H''$ and $r''$ are finite and their finite jumps are allowed.
Indeed, if the metric (\ref{Semiz-I-twopara2}) or (\ref{metric-Buchdahl-v}) is not $C^{1,1}$ but just $C^1$, the divergence of $H''$ or $r''$ leads to a curvature singularity.
(See Lemma~3 in Ref.~\cite{Maeda:2024lbq}.)
We note that a thin shell is described by a $C^{0,1}$ metric, on which $H$ and $r$ are continuous and $H'$ or $r'$ admits a finite jump, so that it is a curvature singularity but the Riemann tensor can still be treated as the Dirac delta function, which is a distribution of order zero~\cite{Steinbauer:2022hvq}.

In fact, by Proposition~2 in Ref.~\cite{Maeda:2024lbq}, $x=x_{\rm h}$ is a curvature singularity unless $\chi\in[-1/3,0)$ or $\chi=-1$.
Similar to the case with $\chi=-1$, solutions with $\chi\in[-1/3,0)$ admit non-degenerate Killing horizons, characterized by $H'(x_{\rm h})\ne 0$, depending on the parameters by Proposition~6 in Ref.~\cite{Maeda:2024lbq} and the spacetime can be extended beyond there.
Then, for $\chi\ne -1/(1+2N)$ with $N\in\mathbb{N}$, the metric is not $C^{[2+\beta],1}$ but $C^{[2+\beta]}$ on the horizon, where $\beta$ is defined by 
\begin{align}
\beta:=-\frac{1+3\chi}{2\chi}. \label{def-beta}
\end{align}
For $\chi= -1/(1+2N)$, or equivalently for $\beta(=N-1)\in\mathbb{N}_0$, the metric can be $C^{\infty}$ on the horizon by Proposition~9 in Ref.~\cite{Maeda:2024lbq}.

\subsection{Matter fields and energy conditions}
\label{sec:ECs}

The standard energy conditions consist of the {\it null} energy condition (NEC), {\it weak} energy condition (WEC), {\it dominant} energy condition (DEC), and {\it strong} energy condition (SEC)~\cite{Maeda:2018hqu}.
While the DEC implies the WEC, both of the WEC and SEC imply the NEC. 
Hence, all the standard energy conditions are violated if the NEC is violated.
In a static region with $H>0$, equivalent representations of those conditions for a perfect fluid obeying $p=\chi\rho$ are given by
\begin{align}
\mbox{NEC}:&~~(1+\chi)\rho\ge 0,\label{NEC-I}\\
\mbox{WEC}:&~~\rho\ge 0\mbox{~in addition to NEC},\label{WEC-I}\\
\mbox{DEC}:&~~(1-\chi)\rho\ge 0\mbox{~in addition to WEC},\label{DEC-I}\\
\mbox{SEC}:&~~(1+3\chi)\rho\ge 0\mbox{~in addition to NEC}.\label{SEC-I}
\end{align}
For $\chi\in[-1/3,0)$, all the standard energy conditions are satisfied (violated) for $\rho\ge(<)0$.

In a region with $H<0$, where $t$ and $x$ are spacelike and timelike coordinates, respectively, the spacetime described by the metric (\ref{Semiz-I-twopara2}) is dynamical and filled not with a perfect fluid but an anisotropic fluid whose energy-momentum tensor is given by 
\begin{align}
\label{Tab-a}
\begin{aligned}
&{T}_{\mu\nu}=({\bar \rho}+{\bar p}_{\rm t}){\bar u}_\mu {\bar u}_\nu+({\bar p}_{\rm r}-{\bar p}_{\rm t})s_\mu s_\nu +{\bar p}_{\rm t}g_{\mu\nu},\\
&{\bar \rho}=-p,\qquad {\bar p}_{\rm r}=-\rho,\qquad {\bar p}_{\rm t}=p,\\
&{\bar u}^\mu\partial_\mu=(-H)^{1/2}\partial_x,\qquad s^\mu\partial_\mu=(-H)^{-1/2}\partial_t,
\end{aligned}
\end{align}
where ${\bar\rho}$, ${\bar p}_{\rm r}$, and ${\bar p}_{\rm t}$ are the energy density, radial pressure, and tangential pressure of the fluid, respectively, and a normalized spacelike vector $s^\mu$ is orthogonal to the normalized fluid element ${\bar u}^\mu$ such that ${\bar u}_\mu {\bar u}^\mu=-1$, $s_\mu s^\mu=1$, and ${\bar u}_\mu s^\mu=0$.

Alternatively, the matter field in a region with $H<0$ may be written as
\begin{align}
\label{T-spacelike}
\begin{aligned}
&T_{\mu\nu}={\tilde\rho} s_\mu s_\nu-{\tilde p}(g_{\mu\nu}-s_\mu s_\nu),\\
&{\tilde \rho}=-\rho, \qquad {\tilde p}=-p,\\
&s^\mu\partial_\mu=(-H)^{-1/2}\partial_t,
\end{aligned}
\end{align}
where $s^\mu$ is a unit spacelike vector satisfying $s_\mu s^\mu$=1.
Since $h_{\mu\nu}=g_{\mu\nu}-s_\mu s_\nu$ is the induced metric on a timelike hypersurface orthogonal to $s^\mu$, the matter field (\ref{T-spacelike}) may be interpreted as a {\it spacelike} (or {\it tachyonic}) perfect fluid with the ``energy density'' ${\tilde \rho}$ and ``pressure'' ${\tilde p}$.
In Appendix~\ref{app:EC-tachyon}, we derive equivalent representations of the standard energy conditions for a tachyonic perfect fluid (\ref{T-spacelike}) in arbitrary $n(\ge 3)$ dimensions without assuming any spacetime symmetry.

Although both matter fields (\ref{Tab-a}) and (\ref{T-spacelike}) give the components (\ref{T-diag}), $\rho$ is not energy density but radial pressure in a region with $H<0$ because $t$ is not a timelike but a spacelike radial coordinate there.
As a consequence, equivalent representations of the standard energy conditions in the dynamical region with $H<0$ are given by 
\begin{align}
\mbox{NEC}:&~~(1+\chi)\rho\le 0,\label{NEC-I-in}\\
\mbox{WEC}:&~~\chi\rho\le 0\mbox{~in addition to NEC},\label{WEC-I-in}\\
\mbox{DEC}:&~~(1-\chi)\rho\ge 0\mbox{~in addition to WEC},\label{DEC-I-in}\\
\mbox{SEC}:&~~(1-\chi)\rho\le 0\mbox{~in addition to NEC}.\label{SEC-I-in}
\end{align}
For $\chi\in[-1/3,0)$, all the standard energy conditions are violated for $\rho>0$.
For $\rho<0$, the NEC and SEC are satisfied, while the WEC is violated (and hence the DEC, too).

If the metric (\ref{metric-Buchdahl-v}) is $C^\infty$ at a non-degenerate Killing horizon, which is possible only for $\chi= -1/(1+2N)$ with $N\in\mathbb{N}$, or equivalently for $\beta(=N-1)\in\mathbb{N}_0$, the asymptotic solution near a non-degenerate Killing horizon $x=x_{\rm h}$ is given by 
\begin{align}
&H(x)\simeq \sum_{i=1}^{3+\beta}H_i\Delta^i,\\
&r(x)\simeq r_0+r_1\Delta+r_{2+\beta}\Delta^{2+\beta},\\
&\rho\simeq \rho_{1+\beta}\Delta^{1+\beta},
\end{align}
where $\Delta:=x-x_{\rm h}$ and 
\begin{align}
&H_1=\frac{1}{r_0r_1},\qquad \rho_{1+\beta}=-\frac{2(2+\beta)(1+\beta)H_1r_{2+\beta}}{\kappa(1+\chi)r_0}.
\end{align}
(See Proposition~9 in Ref.~\cite{Maeda:2024lbq}.)
The coefficient $r_1(\ne 0)$ can be set to one without loss of generality by rescaling transformations of $t$ and $x$.
Parameters of the asymptotic solution are $r_0(> 0)$ and $r_{2+\beta}$ (or $\rho_{1+\beta}$) that determine other coefficients.
This asymptotic solution shows the following proposition.
%----------------------- lemma ------------------------------%
\begin{Prop}
\label{Pro:EC-horizon}
Consider a non-vacuum solution described by the metric (\ref{metric-Buchdahl-v}) to the Einstein equations with a perfect fluid (\ref{EFE}) obeying a linear equation of state $p=\chi\rho$ with $\chi= -1/(1+2N)$ where $N\in\mathbb{N}$ and suppose that the metric is $C^\infty$ at a non-degenerate Killing horizon $x=x_{\rm h}$.
Then, the standard energy conditions satisfied near the horizon are given in the following table.
\begin{center}
\begin{tabular}{|c|c||c|c|c|}\hline\hline
& & $x<x_{\rm h}$ & $x_{\rm h}<x$ \\ \hline
Odd $N$ (even $\beta$) & $\rho_{1+\beta} > 0$ & NEC, SEC & All \\ \cline{2-4}
& $\rho_{1+\beta}< 0$ & None & None \\ \hline
Even $N$ (odd $\beta$) & $\rho_{1+\beta}> 0$ & None & All \\ \cline{2-4}
& $\rho_{1+\beta}< 0$ & NEC, SEC & None \\ \hline\hline
\end{tabular}
\end{center}
\end{Prop}
{\it Proof}. 
If the metric (\ref{metric-Buchdahl-v}) is $C^\infty$ at the horizon, $\rho_{1+\beta}$ takes the same value on both sides of the horizon, so that the sign of $\rho$ is the same across the horizon for odd $\beta$, while it differs on the other side of the horizon for even $\beta$.
Then, the proposition follows from Eqs.~(\ref{NEC-I})--(\ref{SEC-I}) and Eqs.~(\ref{NEC-I-in})--(\ref{SEC-I-in}).
\qed
%----------------------- lemma ------------------------------%

By Proposition~2 in Ref.~\cite{Maeda:2021ukk}, if $r''(x_{\rm h})= 0$ is satisfied, matter fields are absent on the horizon except for $\chi=-1$.
If $r''(x_{\rm h})\ne 0$ is satisfied, there is a matter field of the Hawking-Ellis type II on the horizon.
For $\chi\in[-1/3,0)$, by Proposition~8 in Ref.~\cite{Maeda:2024lbq}, $r''(x_{\rm h})\ne 0$ is satisfied only for $\chi=-1/3$ and the energy-momentum tensor on the horizon is equivalent to a null dust fluid that is written in the single-null coordinates (\ref{metric-Buchdahl-v}) with the upper sign as 
\begin{align}
\label{Tab-horizon}
\begin{aligned}
&{T}_{\mu\nu}=\mu k_\mu k_\nu,\\
&k^\mu\partial_\mu=\partial_v,\qquad \mu=-\frac{2r''}{\kappa r}\biggl|_{x=x_{\rm h}},
\end{aligned}
\end{align}
where $k_\mu k^\mu=0$ holds.
By Proposition~19 in Ref.~\cite{Maeda:2018hqu}, the null dust fluid satisfies (violates) all the standard energy conditions if $\mu\ge (<)0$.

\subsection{Whittaker perfect-fluid solution for $p=-\rho/3$}

The Whittaker solution is an exact static spherically symmetric solution in the system (\ref{EFE}) with an equation of state $\rho=-3p+2\mu_0$, where $\mu_0$ is a constant~\cite{Whittaker}.
This perfect fluid is equivalent to a combination of a perfect fluid obeying $\rho=-3p$ and a cosmological constant $\Lambda(=-\kappa\mu_0)$.
The Whittaker metric without a cosmological constant is given in the conventional coordinates $(T,r,\theta,\phi)$ as
\begin{align}
\label{whittaker}
\begin{aligned}
&\D s^2=-f\D T^2+\frac{\D r^2}{(1-\alpha r^2)f}+r^2\D\Omega^2,\\
&f(r)=1-\frac{2M\sqrt{1-\alpha r^2}}{r},
\end{aligned}
\end{align}
where $M$ and $\alpha$ are parameters.
It solves the Einstein equations (\ref{EFE}) with
\begin{align}
\label{whittaker-matter}
&u^\mu\partial_\mu=f^{-1/2}\partial_T,\qquad \rho=-3p=\frac{3\alpha }{\kappa}f.
\end{align}
For $\alpha=0$, the Whittaker solution (\ref{whittaker}) reduces to the Schwarzschild vacuum solution.
For $M\ne 0$, there is a curvature singularity at the center $r=0$.
For $M=0$ with $\alpha\ne 0$, in contrast, the solution describes a non-singular static universe and its spatial topology is ${\rm S}^3$ for $\alpha>0$ and ${\rm H}^3$ for $\alpha<0$.

The global structure of the Whittaker spacetime was studied in~\cite{Cho:2016kpf}.
In this paper, we focus on the parameter region given by 
\begin{align}
M>0,\qquad \alpha>-\frac{1}{4M^2}.
\end{align}
Then, the Whittaker solution admits a Killing horizon at 
\begin{align}
r=r_{\rm h}=\frac{2M}{\sqrt{1+4\alpha M^2}}(>0),\label{rh-W}
\end{align}
which satisfies $f(r_{\rm h})=0$ and $1-\alpha r_{\rm h}^2>0$.
The coordinates (\ref{whittaker}) cover the domains $r\in (0,r_{\rm h})$ and $r\in (r_{\rm h},1/\sqrt{\alpha})$ for $\alpha>0$ and $r\in (0,r_{\rm h})$ and $r\in (r_{\rm h},\infty)$ for $\alpha< 0$.

\subsubsection{Comoving quasi-global coordinates}
\label{sec:quasi-global-W}

Here we present the Whittaker solution in the comoving quasi-global coordinates (\ref{Semiz-I-twopara2}).
We introduce the quasi-global coordinates $(t,x)$ as
\begin{align}
\label{trans+}
t:=\sqrt{\omega}T,\qquad x:=\frac{1}{\sqrt{\omega}}\int\frac{\D r}{\sqrt{1-\alpha r^2}},
\end{align}
where $\omega$ is a positive constant to be determined as a gauge choice.
Then, we obtain
\begin{align}
\label{matter-x}
\begin{aligned}
&H(x):=\omega^{-1}f(r(x)),\\
&u^\mu\partial_\mu=\sqrt{\omega}H^{-1/2}\partial_t,\\
&\rho=-3p=\frac{3\alpha\omega}{\kappa}H.
\end{aligned} 
\end{align} 
The Killing horizon (\ref{rh-W}) corresponds to $x=x_{\rm h}$ determined by $H(x_{\rm h})=0$.

For $\alpha>0$, Eq.~(\ref{trans+}) gives 
\begin{align}
\label{r+}
\begin{aligned}
x=\frac{1}{\sqrt{\alpha\omega}}\arcsin(\sqrt{\alpha}r)(\ge 0)~~\Leftrightarrow~~&r(x)=\frac{1}{\sqrt{\alpha}}\sin(\sqrt{\alpha\omega}x).
\end{aligned}
\end{align} 
The other metric function is given by 
\begin{align}
\label{Hr+}
&H(x)=\frac{1}{\omega}\biggl(1-\frac{2\sqrt{\alpha}M}{\tan(\sqrt{\alpha\omega}x)}\biggl),
\end{align} 
which shows
\begin{align}
x_{\rm h}=\frac{1}{\sqrt{\alpha\omega}}\arctan(2\sqrt{\alpha}M)\biggl(<\frac{\pi}{2\sqrt{\alpha\omega}}\biggl).
\end{align} 
The domain of $x$ in the single-null coordinates (\ref{metric-Buchdahl-v}) is $0<x<\pi/\sqrt{\alpha\omega}$ and both of the boundaries $x=0$ and $x=\pi/\sqrt{\alpha\omega}$ correspond to the singularity $r=0$.
On the other hand, $r=r_{\rm b}:=1/\sqrt{\alpha}$ for $\alpha>0$ corresponds to $x=x_{\rm b}:=\pi/(2\sqrt{\alpha\omega})$.

For $-1/(4M^2)<\alpha<0$, Eq.~(\ref{trans+}) gives 
\begin{align}
\label{r-}
\begin{aligned}
x=\frac{1}{\sqrt{|\alpha|\omega}}\mbox{arcsinh}(\sqrt{|\alpha|}r)(\ge 0)~~\Leftrightarrow~~r(x)=\frac{1}{\sqrt{|\alpha|}}\sinh(\sqrt{|\alpha|\omega}x).
\end{aligned} 
\end{align} 
The other metric function is given by 
\begin{align}
\label{Hr-}
H(x)=&\frac{1}{\omega}\biggl(1-\frac{2\sqrt{|\alpha|}M}{\tanh(\sqrt{|\alpha|\omega}x)}\biggl),
\end{align} 
which shows
\begin{align}
x_{\rm h}=\frac{1}{\sqrt{|\alpha|\omega}}{\rm arctanh}(2\sqrt{|\alpha|}M).\label{xh-W-}
\end{align} 
The domain of $x$ in the single-null coordinates (\ref{metric-Buchdahl-v}) is $0<x<\infty$ and $x=0$ corresponds to the singularity $r=0$.

\subsubsection{Global structure}

Here we study boundaries of the Whittaker spacetime in the quasi-global coordinates (\ref{Semiz-I-twopara2}).
Let $\gamma$ be a radial geodesic parametrized by an affine parameter $\lambda$ as $x^\mu=(t(\lambda),x(\lambda),0,0)$ with its tangent vector $k^\mu={\dot x}^\mu$, where a dot denotes differentiation with respect to $\lambda$.
Along $\gamma$, $E:=-\xi_\mu k^\mu=H{\dot t}$ is conserved, where $\xi^\mu=(1,0,0,0)$ is a hypersurface orthogonal Killing vector.
Combined with the normalization condition $k_\mu k^\mu=\varepsilon$, where $\varepsilon=1,0,-1$ corresponds to the case where $\gamma$ is spacelike, null, and timelike, respectively, we obtain 
\begin{align}
\label{geodesic}
{\dot x}^2=\varepsilon H+E^2~~\to~~\lambda=\pm\int\frac{\D x}{\sqrt{\varepsilon H+E^2}},
\end{align}
where the sign on the right-hand side is chosen to decide the direction of $\gamma$.
The above expression shows $\lim_{x\to \infty}|\lambda|\to \infty$ along null geodesics ($\varepsilon=0$) and therefore $x\to \infty$ is null infinity.
On the other hand, a hypersurface given by a finite value of $x$ is extendible along null geodesics if it is regular.

The causal nature of boundaries of the Whittaker spacetime (\ref{whittaker}) is determined by the two-dimensional Lorentzian portion that is written in the conformally flat form as
\begin{align}
\label{whittaker-2dim}
\begin{aligned}
&\D s_2^2=H(x(x_*))(-\D t^2+\D x_*^2),\\
&x_*:=\int\frac{\D x}{H(x)}.
\end{aligned}
\end{align}
A radius $x=x_0$ is causally null in the Penrose diagram if $\lim_{x\to x_0}|x_*|\to \infty$, while it is causally timelike (spacelike) if it corresponds to a finite value of $x_*$ with $H>(<)0$.

Near $x=x_{\rm b}$ corresponding to $r=r_{\rm b}(=1/\sqrt{\alpha})$ for $\alpha>0$, we obtain
\begin{align}
H(x)\simeq \frac{1}{\omega}+\frac{2\alpha M}{\sqrt{\omega}}(x-x_{\rm b})+O((x-x_{\rm b})^3),
\end{align} 
which shows $\lim_{x\to x_{\rm b}}|x_*|<\infty$ and $H(x_{\rm b})>0$.
Hence, $x=x_{\rm b}$ is a timelike hypersurface.
On the other hand, near $x\to \infty$ ($r\to \infty$) for $-1/(4M^2)<\alpha<0$, Eq.~(\ref{whittaker-2dim}) shows
\begin{align}
&\lim_{x\to \infty}|x_*|=\lim_{r\to \infty}\biggl|\int\frac{1}{f(r)\sqrt{\omega(1-\alpha r^2)}}\D r\biggl|\propto \lim_{r\to \infty}\int\frac{\D r}{r}\to \infty.
\end{align}
Hence, the null infinity $x\to \infty$ is causally null.
The Whittaker spacetime with $-1/(4M^2)<\alpha<0$ is not asymptotically flat because $R^{\mu\nu}_{~~\rho\sigma}$ does not converge to zero as $x\to \infty$.

Lastly, we study the singularity at $r=0$, which corresponds to $x=0$ for $\alpha\ne 0$ and $x=\pi/\sqrt{\alpha\omega}$ for $\alpha>0$.
Near $x=0$, we obtain
\begin{align}
H(x)\simeq -\frac{2M}{\omega^{3/2}x}
\end{align}
both for $\alpha>0$ and $-1/(4M^2)<\alpha<0$, which shows $\lim_{x\to 0}|x_*|<\infty$ and $H<0$ holds in the region of $0<x<x_{\rm h}$.
Hence, the singularity $x=0$ is spacelike.
On the other hand, near the singularity $x=\pi/\sqrt{\alpha\omega}$ for $\alpha>0$, we obtain
\begin{align}
H(x)\simeq -\frac{2M}{\omega^{3/2}}\biggl(x-\frac{\pi}{\sqrt{\alpha\omega}}\biggl)^{-1},
\end{align}
which shows $\lim_{x\to 0}|x_*|<\infty$ and $H>0$ holds in the region of $x_{\rm h}<x<\pi/\sqrt{\alpha\omega}$.
Hence, the singularity $x=\pi/\sqrt{\alpha\omega}$ is timelike.

\begin{figure}[htbp]
\begin{center}
\includegraphics[width=0.55\linewidth]{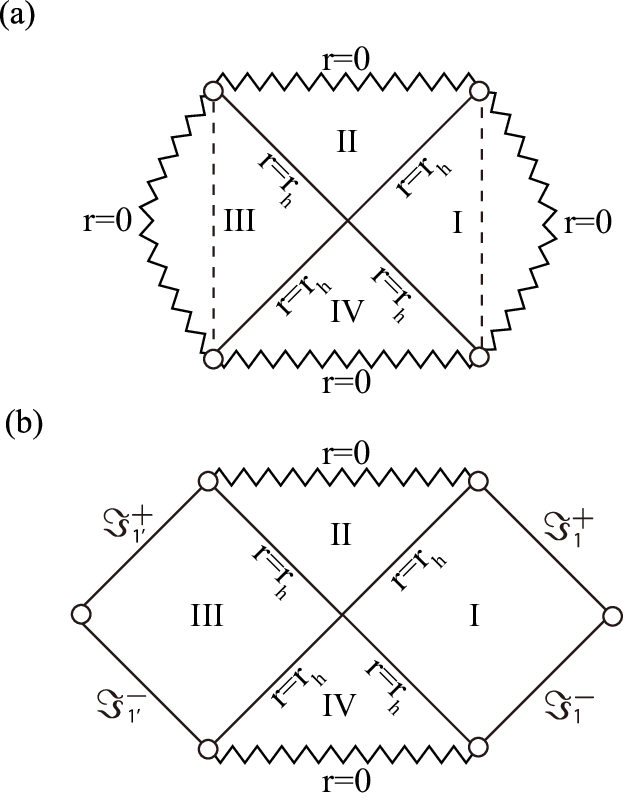}
\caption{\label{Fig-PenroseDiagrams-Whittaker} Penrose diagrams of the maximally extended Whittaker spacetime with $M>0$ for (a) $\alpha>0$ and (b) $-1/(4M^2)<\alpha<0$ in the case where the values of $\alpha$ and $M$ in the regions I, II, III, and IV are the same. A zig-zag line is the curvature singularity at $r=0$. A dashed line in (a) corresponds to $r=r_{\rm b}$. The symbol $\Im^{+(-)}$ in (b) stands for a future (past) null infinity. Note that the Killing horizon $r=r_{\rm h}$ in (a) is not an event horizon due to the absence of null infinity.}
\end{center}
\end{figure}

The extension beyond the Killing horizon $x=x_{\rm h}$ is possible in the single-null coordinates (\ref{metric-Buchdahl-v}).
In the case where the values of $M$ and $\alpha$ are the same in the whole spacetime, the metric is analytic ($C^\omega$) on the horizon, and possible Penrose diagrams of the maximally extended Whittaker spacetime are shown in Fig.\ref{Fig-PenroseDiagrams-Whittaker}.
%------------<fig>---------------------------
%--------------<fig>-----------------------

%======================================%
%<<<<<<<<<<<< SECTION I >>>>>>>>>>>>>>%
%======================================%
\section{Fake Schwarzschild black holes}
\label{sec:main}

In this section, we present two different solutions describing a fake Schwarzschild black hole.

\subsection{Semiz interior for $\chi=-1/5$}
\label{sec:Semiz}
First, we use Semiz's solution for $\chi=-1/5$~\cite{Semiz:2020lxj}, with which Eq.~(\ref{def-beta}) gives $\beta=1$.
This construction is a particular four-dimensional case among the results in Ref.~\cite{Maeda:2022lsm} in four and five dimensions, in which the solution is referred to as {\it the Semiz class-I solution}.
The Semiz class-I solution is defined in arbitrary $n(\ge 4)$ dimensions for $\chi=-(n-3)/(n+1)$ and it admits Killing horizons only for $n=4$ and $5$ as it becomes a curvature singularity for $n\ge 6$.

The Semiz class-I solution in four dimensions is given in the comoving quasi-global coordinates (\ref{Semiz-I-twopara2}) as 
\begin{align}
\begin{aligned}
&H(x)=\frac{x-2M}{x-{\zeta}\left(x-2M\right)^{3}},\\
&r(x)=x-{\zeta}\left(x-2M\right)^3,\\
&\rho=-5p=\frac{15{\zeta}}{\kappa}H^2.
\end{aligned}
\end{align} 
The solution is parametrized by $M$ and $\zeta$ and reduces to the Schwarzschild vacuum solution with the ADM mass $M$ for ${\zeta}\to 0$.
The spacetime is asymptotically flat as $x\to \infty$ and admits a single non-degenerate Killing horizon at $x=x_{\rm h}(=2M)$ for $M>0$.

In a static region with $H>0$, the perfect fluid satisfies (violates) all the standard energy conditions for $\zeta>(<)0$.
In a dynamical region with $H<0$, the NEC and SEC are satisfied but the WEC is violated for $\zeta<0$.
In contrast, all the standard energy conditions are violated for $\zeta>0$.
On the Killing horizon $x=x_{\rm h}$, matter fields are absent as $r''(x_{\rm h})=0$ holds.

\begin{figure}[htbp]
\begin{center}
\includegraphics[width=0.55\linewidth]{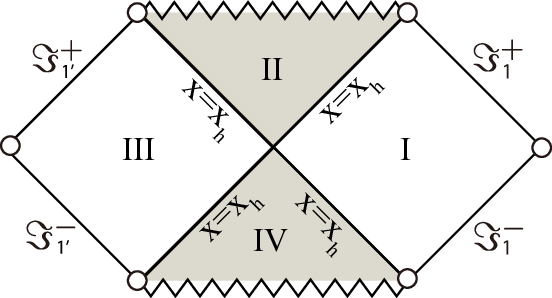}
\caption{\label{Fig-PenroseDiagrams-FakeSemiz}A Penrose diagram of a fake Schwarzschild black hole with a Semiz dynamical interior (shaded). The metric is $C^{2,1}$ on the Killing horizon $x=x_{\rm h}(=2M_+)$. }
\end{center}
\end{figure}

A fake Schwarzschild black hole can be constructed under the NEC and SEC by attaching the Schwarzschild exterior spacetime with $\zeta=0$ and $M=M_+(>0)$ defined in the domain $x>x_{\rm h}$ at the Killing horizon $x=x_{\rm h}$ to the Semiz interior (dynamical) spacetime with $\zeta<0$ and $M=M_+$ defined in the domain $x<x_{\rm h}$.
Then, by the following expressions 
\begin{align}
\begin{aligned}
&H(2M_+)=0,\qquad r(2M_+)=2M_+,\\
&H'(2M_+)=\frac{1}{2M_+},\qquad r'(2M_+)=1,\\
&H''(2M_+)=-\frac{1}{2M_+^2},\qquad r''(2M_+)=0,\\
&H'''(2M_+)=\frac{3}{4M_+^3},\qquad r'''(2M_+)=-6\zeta,
\end{aligned} 
\end{align} 
the metric in the single-null coordinates (\ref{metric-Buchdahl-v}) is $C^{2,1}$ on the horizon $x=2M_+$ and therefore there is no lightlike massive thin shell there.
For $\zeta<0$, $r(x)$ is a monotonically increasing function of $x$ shown by $r'=1-3\zeta(x-2M)^2>0$, so that there is a curvature singularity at $x=x_{\rm s}(<2M_+)$ inside the Killing horizon determined by $r(x_{\rm s})=0$.
The Penrose diagram of the resultant fake Schwarzschild black hole is shown in Fig.~\ref{Fig-PenroseDiagrams-FakeSemiz}\footnote{The Penrose diagram of the fake Schwarzschild black hole can be different if we use the Semiz interior solution with $\zeta>0$ and $M=M_+(>0)$ in the domain $x<x_{\rm s}$, where all the standard energy conditions are violated.
In this case, while the Penrose diagram is the same as Fig.~\ref{Fig-PenroseDiagrams-FakeSemiz} for $0<M_+\le M_{\rm ex(4+)}$, where $M_{\rm ex(4+)}:=1/(3\sqrt{3\zeta})$, the spacelike singularity in Fig.~\ref{Fig-PenroseDiagrams-FakeSemiz} is replaced by a big bounce for $M_+>M_{\rm ex(4+)}$, so that the resulting spacetime describes the so-called ``black bounce'' shown in Fig.~3(b) in Ref.~\cite{Maeda:2022lsm}.}.
%------------<fig>---------------------------
%--------------<fig>-----------------------

\begin{figure}[htbp]
\begin{center}
\includegraphics[width=0.55\linewidth]{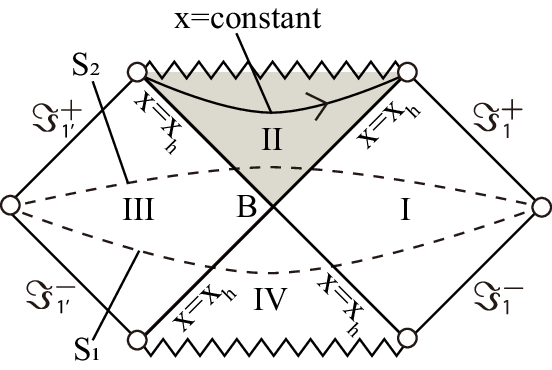}
\caption{\label{Fig-PenroseDiagrams-FakeSemiz2}A Penrose diagram of a fake Schwarzschild black hole with a Semiz dynamical interior only in the region II (shaded). A curve in the region II is the orbit of a fluid element of a spacelike perfect fluid (\ref{T-spacelike}).}
\end{center}
\end{figure}

We note that the values of $\zeta$ in the Semiz regions II and IV may differ.
Hence, the Penrose diagram of the simplest fake Schwarzschild black hole is the one shown in Fig.~\ref{Fig-PenroseDiagrams-FakeSemiz2}.
It is observed that there is no matter field on the Cauchy surface S${}_1$ in the past, but there is on the Cauchy surface S${}_2$ in the future.
Although such a counterintuitive configuration could suggest that the bifurcation two-sphere (B) is singular, it is not the case as shown below.
%------------<fig>---------------------------
%--------------<fig>-----------------------

For static spherically symmetric solutions with $\chi\in[-1/3,0)$, the asymptotic behaviors of the metric functions near a non-degenerate Killing horizon $x=x_{\rm h}$ are given by 
\begin{align}
\begin{aligned}
&H(x)\simeq \sum_{i=1}^{3+\beta}H_i\Delta^i,\\
&r(x)\simeq r_0+r_1\Delta+r_{2+\beta}\Delta^{2+\beta},
\end{aligned} 
\end{align} 
where $\Delta:=x-x_{\rm h}$ and $\beta$ is defined by Eq.~(\ref{def-beta}).
(See the proof of Proposition~6 in Ref.~\cite{Maeda:2024lbq}.)
Here $H_1$, $H_2$, $r_0$, and $r_1$ are non-zero in particular.
Using the tortoise coordinate $x_*:=\int H(x)^{-1}\D x$, we define the null Kruskal-Szekeres coordinates $U$ and $V$ by
\begin{align}
U:=\mp e^{-H_1u/2},\qquad V:=e^{H_1v/2},\label{def-UV}
\end{align} 
where $u:=t-x_*$ and $v:=t+x_*$.
For a non-degenerate outer Killing horizon, characterized by $H_1>0$, $x=x_{\rm h}$ corresponds to $x_*\to -\infty$ and $UV= 0$.
The upper (lower) sign in the definition of $U$ in Eq.~(\ref{def-UV}) corresponds to $x>(<)x_{\rm h}$.
In the null Kruskal-Szekeres coordinates, the metric (\ref{Semiz-I-twopara2}) is written as
\begin{align}
\label{Kruskal}
\begin{aligned}
&\D s^2=2g_{UV}\D U\D V+r(x(U,V))^2\D\Omega^2,\\
&g_{UV}=\frac{2H(x(U,V))}{H_1^2UV}.
\end{aligned} 
\end{align} 
The asymptotic behaviors of the metric functions near a Killing horizon $x_*\to -\infty$ ($UV=0$) are given by 
\begin{align}
\label{expansion-UV}
\begin{aligned}
g_{UV}\simeq& \frac{2}{H_1^2}\biggl(-1+2\frac{H_2}{H_1^2}UV -3\frac{H_2^2}{H_1^4}U^2V^2+O(U^3V^3)\biggl),\\
r^2\simeq& r_0^2-2\frac{r_0r_1}{H_1}UV+O(U^2V^2).
\end{aligned} 
\end{align} 
Equation~(\ref{expansion-UV}) shows that a bifurcation two-sphere given by $U=V=0$ is regular if $H'(x_{\rm h})$, $H''(x_{\rm h})$, $r(x_{\rm h})$, and $r'(x_{\rm h})$ are the same on both sides of the horizon so that the metric and its inverse in the coordinates (\ref{Kruskal}) are at least $C^{1,1}$ there.

Thus, the counterintuitive configuration shown in Fig.~\ref{Fig-PenroseDiagrams-FakeSemiz2} could be attributed to the violation of either the WEC or DEC.
Because a fluid element of the spacelike (tachyonic) perfect fluid in the region II moves in a spacelike direction with its tangent vector $s^\mu$ in Eq.~(\ref{T-spacelike}), it does not cross a Killing horizon nor a bifurcation two-sphere, which is analogous to the fact that a fluid element of a (timelike) perfect fluid in a static region with $H>0$ never crosses a horizon nor a bifurcation two-sphere.

In fact, such a configuration of black holes contradicts the {\it conservation theorem} resulting from the DEC.
This theorem asserts that if $T_{\mu\nu}=0$ holds on a closed achronal hypersurface $\Sigma$, then $T_{\mu\nu}=0$ holds everywhere in the domain of dependence of $\Sigma$~\cite{Hawking:1973uf}.
(See also Proposition~2.2.3 in Ref.~\cite{Iizuka:2025xnd}.)
Since we can prepare a closed achronal hypersurface $\Sigma$ with $T_{\mu\nu}=0$ that spans the regions I, IV, and III and contains the region IV in its future domain of dependence in Fig.~\ref{Fig-PenroseDiagrams-FakeSemiz2}, the DEC must be violated at least somewhere in the region IV by the contrapositive of the conservation theorem.

\subsection{Whittaker interior for $\chi=-1/3$}
\label{sec:extension}

Next, we use the Whittaker solution (\ref{whittaker}) for $\chi=-1/3$, with which Eq.~(\ref{def-beta}) gives $\beta=0$.
The solution is written in the comoving quasi-global coordinates in Sec.~\ref{sec:quasi-global-W}.
By Eqs.~(\ref{trans+}) and (\ref{r+})--(\ref{xh-W-}), the derivatives of the metric functions are computed to give
\begin{align}
\begin{aligned}
&H'(x)=\frac{2M}{\sqrt{\omega}r^2},\qquad H''(x)=-\frac{4M}{r^3}\sqrt{1-\alpha r^2},\\
&r'(x)=\sqrt{\omega(1-\alpha r^2)},\qquad r''(x)=-\omega\alpha r,
\end{aligned} 
\end{align} 
of which the values on the Killing horizon are
\begin{align}
\begin{aligned}
&H'(x_{\rm h})=\frac{2M}{\sqrt{\omega}r_{\rm h}^2},\qquad H''(x_{\rm h})=-\frac{4M}{r_{\rm h}^3}\sqrt{1-\alpha r_{\rm h}^2},\\
&r'(x_{\rm h})=\sqrt{\omega(1-\alpha r_{\rm h}^2)},\qquad r''(x_{\rm h})=-\omega\alpha r_{\rm h}.
\end{aligned} 
\end{align} 
Now we choose the gauge constant $\omega$ such that $r'(x_{\rm h})=1$, namely
\begin{align}
\omega=\frac{1}{1-\alpha r_{\rm h}^2}=1+4\alpha M^2(>0).
\end{align} 
Then, we obtain
\begin{align}
\label{ddg-h}
\begin{aligned}
&H'(x_{\rm h})=\frac{\sqrt{1+4\alpha M^2}}{2M}=r_{\rm h}^{-1},\qquad H''(x_{\rm h})=-\frac{4M}{r_{\rm h}^3\sqrt{1+4\alpha M^2}}=-\frac{2}{r_{\rm h}^2},\\
&r'(x_{\rm h})=1,\qquad r''(x_{\rm h})=-2\alpha M\sqrt{1+4\alpha M^2}=-\frac{4\alpha M^2}{r_{\rm h}}.
\end{aligned} 
\end{align} 
A matter field (\ref{Tab-horizon}) exists on the horizon for $\alpha M \ne 0$ as $r''|_{x=x_{\rm h}}\ne 0$ holds.
It satisfies (violates) all the standard energy conditions for $\alpha>(<)0$.

Now we consider two Whittaker spacetimes $({\cal M}_+^4,g^+_{\mu\nu})$ and $({\cal M}_-^4,g^-_{\mu\nu})$ to study non-analytic extensions beyond a Killing horizon.
Let $x_\pm=x_{\rm h(\pm)}$ be the location of the horizon determined by $H_\pm(x_{\rm h(\pm)})=0$ in the Whittaker spacetime $({\cal M}_\pm^4,g^\pm_{\mu\nu})$ with the parameters $(\alpha,M)=(\alpha_\pm,M_\pm)$ described in the single-null coordinates $(v_\pm,x_\pm,\theta,\phi)$ as
\begin{equation}
\D s^2=-H_\pm(x_\pm)\D v_\pm^2+2\D v_\pm\D x_\pm+r_\pm(x_\pm)^2\D\Omega^2.\label{metric-Buchdahl-v+}
\end{equation}
We note that $x_{\rm h(+)}\ne x_{\rm h(-)}$ holds in general.
We glue an ``outer'' Whittaker solution $({\cal M}_+^4,g^+_{\mu\nu})$ defined in the domain $x_+>x_{\rm h(+)}$ to an ``inner'' Whittaker solution $({\cal M}_-^4,g^-_{\mu\nu})$ defined in the domain $x_-<x_{\rm h(-)}$ at the Killing horizon $\Sigma$, which is a regular null hypersurface $x_\pm=x_{\rm h(\pm)}$ in $({\cal M}_\pm^4,g^\pm_{\mu\nu})$.
As the first junction conditions, the induced metric must be continuous on the horizon, which requires $r_+(x_{\rm h(+)})=r_-(x_{\rm h(-)})(=r_{\rm h})$ and therefore
\begin{align}
\frac{M_+}{\sqrt{1+4\alpha_+ M_+^2}}=\frac{M_-}{\sqrt{1+4\alpha_- M_-^2}}.\label{condition-key}
\end{align}
We note that $\alpha_+ M_+^2=\alpha_+ M_+^2$ is equivalent to $(\alpha_+,M_+)=(\alpha_-,M_-)$. 
Hence, under the condition (\ref{condition-key}), the Killing horizon is regular as Eq.~(\ref{ddg-h}) shows that the metric is analytic ($C^\omega$) on the horizon for $(\alpha_+,M_+)=(\alpha_-,M_-)$ and $C^{1,1}$ otherwise.

\begin{figure}[htbp]
\begin{center}
\includegraphics[width=0.55\linewidth]{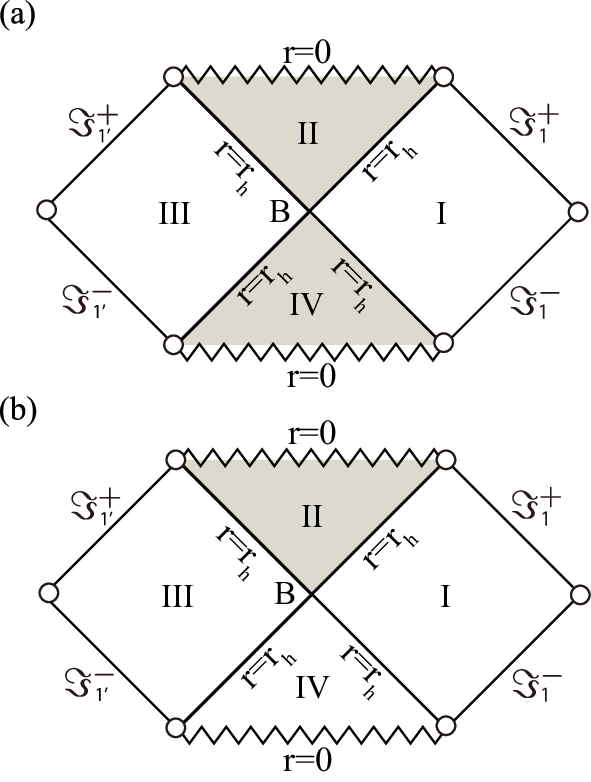}
\caption{\label{Fig-PenroseDiagrams-Whittaker-FakeBH} Penrose diagrams of a fake Schwarzschild black hole with a Whittaker dynamical interior (shaded) (a) in the regions II and IV and (b) only in the region II. The metric in the single-null coordinates (\ref{metric-Buchdahl-v+}) is $C^{1,1}$ on the Killing horizon between the Schwarzschild and Whittaker regions.}
\end{center}
\end{figure}

As a special case, a Schwarzschild exterior for $(\alpha,M)=(0,M_+)$ and a Whittaker interior for $(\alpha,M)=(\alpha_-,M_-)$ can be attached if $M_+$ satisfies
\begin{align}
M_+=\frac{M_-}{\sqrt{1+4\alpha_-M_-^2}}(>0). \label{condition-fake-W}
\end{align}
The resultant spacetime describes a fake Schwarzschild black hole, of which Penrose diagrams are shown in Fig.~\ref{Fig-PenroseDiagrams-Whittaker-FakeBH}.
Note that the values of $\alpha$ and $M$ in the regions II and IV in Fig.~\ref{Fig-PenroseDiagrams-Whittaker-FakeBH}(a) may differ because the condition (\ref{condition-fake-W}) allows different sets of $(\alpha_-,M_-)$ for a given value of $M_+$.
For $\alpha_->0$, the matter field in the interior region satisfies the NEC and SEC but violates the WEC.
For $-1/(4M^2)<\alpha_-<0$, it violates all the standard energy conditions.
%------------<fig>---------------------------
%--------------<fig>-----------------------

\begin{figure}[htbp]
\begin{center}
\includegraphics[width=0.55\linewidth]{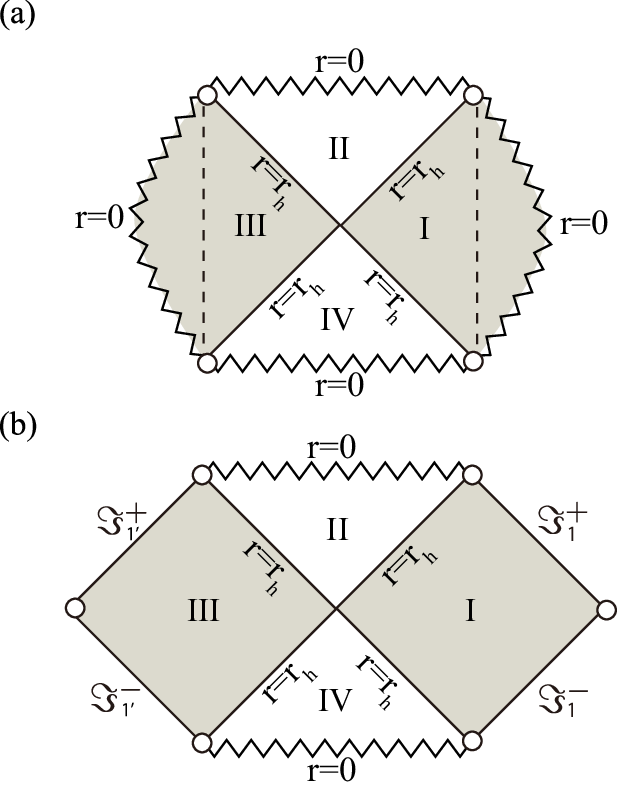}
\caption{\label{Fig-PenroseDiagrams-Whittaker-FakeBH2} Penrose diagrams of a Schwarzschild interior spacetime attached on the Killing horizon to a Whittaker exterior spacetime (shaded) in the case where the values of $\alpha$ and $M$ are the same in the regions I and III and satisfy (a) $\alpha>0$ and (b) $-1/(4M^2)<\alpha<0$. While the Killing horizon is an event horizon in (b), it is not in (a) due to the absence of null infinity.}
\end{center}
\end{figure}

On the contrary, a Whittaker exterior for $(\alpha,M)=(\alpha_+,M_+)$ and a Schwarzschild interior for $(\alpha,M)=(0,M_-)$ can be attached on the horizon if $M_-$ satisfies
\begin{align}
M_-=\frac{M_+}{\sqrt{1+4\alpha_+M_+^2}}(>0).
\end{align}
The resultant spacetime describes a static perfect fluid hovering outside a Schwarzschild black hole, of which Penrose diagrams are shown in Fig.~\ref{Fig-PenroseDiagrams-Whittaker-FakeBH2}.
The perfect fluid satisfies (violates) all the standard energy conditions for $\alpha_+>(<)0$.
Although we assume in Fig.~\ref{Fig-PenroseDiagrams-Whittaker-FakeBH2} that the values of $\alpha$ and $M$ are the same in the regions I and III, they may differ.
%------------<fig>---------------------------
%--------------<fig>-----------------------

The present result confirms Proposition~6 in Ref.~\cite{Maeda:2024lbq} in the special case $\chi=-1/3$.
In Appendix~\ref{app:topo}, we demonstrate the same analysis for the topological generalization of the Whittaker solution in the presence of $\Lambda$.

%======================================%
%<<<<<<<<<<<< SECTION I >>>>>>>>>>>>>>%
%======================================%
\section{Fake Kerr black holes}
\label{sec:main2}

The previous section studied models of fake Schwarzschild black holes with interior regions described by the Semiz class-I and Whittaker solutions that violate the WEC (and the DEC, too).
Here we present a model of rotating {\it fake Kerr} black holes described by the G\"urses-G\"ursey metric.
The interior matter field in this model differs from Eq.~(\ref{T-diag}) and violates only the DEC.

\subsection{G\"urses-G\"ursey metric}

The G\"urses-G\"ursey metric~\cite{Gurses:1975vu} describes a stationary and axisymmetric spacetime and is given in the Boyer-Lindquist coordinates as 
\begin{align}
\label{BL}
\begin{aligned}
\D s^2=&-\biggl(1-\frac{2m(r)r}{\Sigma(r,\theta)}\biggl)\D t^2-\frac{4am(r)r\sin^2\theta}{\Sigma(r,\theta)}\D t\D\phi \\
&+\frac{\Sigma(r,\theta)}{\Delta(r)}\D r^2+\Sigma(r,\theta)\D\theta^2+\biggl(r^2+a^2+\frac{2a^2m(r)r\sin^2\theta}{\Sigma(r,\theta)}\biggl)\sin^2\theta\D\phi^2, \\
&\Sigma(r,\theta):=r^2+a^2\cos^2\theta,\qquad \Delta(r):=r^2+a^2-2rm(r).
\end{aligned}
\end{align} 
Here $m(r)$ is an arbitrary function and $a$ is a constant characterizing the angular momentum of the spacetime.
A regular null hypersurface $r=r_{\rm h}$ determined by $\Delta(r_{\rm h})=0$ is a Killing horizon associated with a Killing vector $\xi^\mu=(1,0,0,a/(r_{\rm h}^2+a^2))$.
With $a=0$, the metric (\ref{BL}) reduces to 
\begin{align}
\label{metric-app}
\D s^2=&-\biggl(1-\frac{2m(r)}{r}\biggl)\D t^2+\biggl(1-\frac{2m(r)}{r}\biggl)^{-1}\D r^2+r^2\D\Omega^2.
\end{align} 
This spherically symmetric metric was introduced in Ref.~\cite{Ovalle:2024wtv} to describe a fake Schwarzschild black hole with an integrable singularity inside.
The mass function $m(r)$ in the metric (\ref{metric-app}) is identical to the Misner-Sharp mass~\cite{Misner:1964je} when we set $G=1$, which is known to be a proper quasi-local mass for spherically symmetric spacetimes in general relativity~\cite{Hayward:1994bu}.
In the rotating case (\ref{BL}), $m(r)$ may also be considered as a natural quasi-local mass because it coincides with the Arnowitt-Deser-Misner (ADM) mass in vacuum and converges to the ADM mass at spacelike infinity in the asymptotically flat non-vacuum spacetime.
The G\"urses-G\"ursey metric (\ref{BL}) and its non-rotating version (\ref{metric-app}) have been used very frequently to construct models of non-singular black holes.
(See Sec.~2.2 in Ref.~\cite{Maeda:2021jdc} for a review.)

We introduce orthonormal basis one-forms $\{E^{(a)}_\mu\}~(a=0,1,2,3)$ given by 
\begin{align}
\label{basis-rotating}
\begin{aligned}
&E^{(0)}_\mu\D x^\mu = \left\{
\begin{array}{ll}
-\sqrt{\Delta/\Sigma}(\D t-a \sin^2\theta\D\phi) & (\mbox{if}~\Delta>0)\\
-\sqrt{-\Sigma/\Delta}\D r & (\mbox{if}~\Delta<0)
\end{array}
\right.,\\
&E^{(1)}_\mu\D x^\mu = \left\{
\begin{array}{ll}
-\sqrt{\Sigma/\Delta}\D r & (\mbox{if}~\Delta>0)\\
-\sqrt{-\Delta/\Sigma}(\D t-a \sin^2\theta\D\phi) & (\mbox{if}~\Delta<0)
\end{array}
\right.,\\
&E^{(2)}_\mu\D x^\mu=\sqrt{\Sigma}\D \theta,\qquad E^{(3)}_\mu\D x^\mu=\frac{\sin\theta}{\sqrt{\Sigma}}\left\{-a\D t+(r^2+a^2)\D\phi\right\},
\end{aligned}
\end{align}
which satisfy $\eta^{(a)(b)}=g^{\mu\nu}{E}_\mu^{(a)} {E}_\nu^{(b)}=\mbox{diag}(-1,1,1,1)$.
Then, orthonormal components of the energy-momentum tensor $T^{\mu\nu}(:=\kappa^{-1}G^{\mu\nu})$ for the corresponding matter field are given by 
\begin{align}
\label{matter-rot}
\begin{aligned}
&{T}^{(a)(b)}:=T^{\mu\nu} {E}_\mu^{(a)} {E}_\nu^{(b)}=\mbox{diag}(\rho,p_1,p_2,p_3),\\
&\rho=-p_1=\frac{2r^2m'}{\kappa\Sigma^2},\qquad p_2=p_3=-\frac{rm''\Sigma+2m'a^2\cos^2\theta}{\kappa\Sigma^2} 
\end{aligned}
\end{align}
independent of the sign of $\Delta$~\cite{Gurses:1975vu,Burinskii:2001bq,Maeda:2021jdc}.

In the Boyer-Lindquist coordinates (\ref{BL}), a Killing horizon $r=r_{\rm h}$ is a coordinate singularity, where $r_{\rm h}m(r_{\rm h})>0$ holds.
A region with $rm(r)\ge 0$ in the G\"urses-G\"ursey spacetime (\ref{BL}) can be expressed in the following Doran coordinates $(\eta,r,\theta,\varphi)$~\cite{Doran:1999gb,Visser:2007fj,Maeda:2021jdc} as
\begin{align}
\D s^2=&-\D \eta^2+\Sigma(r,\theta)\D \theta^2+(r^2+a^2)\sin^2\theta\D\varphi^2\nonumber \\
&+\frac{\Sigma(r,\theta)}{r^2+a^2}\biggl\{\D r+\frac{\sqrt{2m(r)r(r^2+a^2)}}{\Sigma(r,\theta)}(\D \eta-a\sin^2\theta\D\varphi)\biggl\}^2.\label{Doran}
\end{align}
The metric in the Doran coordinates is obtained by the following transformations
\begin{align}
&\D {t}=\D\eta-\frac{\sqrt{2m(r)r(r^2+a^2)}}{\Delta(r)}\D r,\\
&\D{\phi}=\D\varphi-\frac{a}{\Delta(r)}\sqrt{\frac{2m(r)r}{r^2+a^2}}\D r.
\end{align}
Different from the Boyer-Lindquist coordinates, Killing horizons are not coordinate singularities in the Doran coordinates (\ref{Doran}) and $g^{\eta\eta}=-1$ holds.
The spacetime described by the metric (\ref{Doran}) is of Petrov type D everywhere including Killing horizons.
By Corollary~4 in Ref.~\cite{Maeda:2022vld}, the form of the energy-momentum tensor (\ref{matter-rot}) remains valid on Killing horizons.

\subsection{Violation of the dominant energy condition}
\label{sec:violation}

With $m(r)=M$, where $M$ is a constant, the G\"urses-G\"ursey metric (\ref{BL}) reduces to the Kerr metric in the Boyer-Lindquist coordinates.
We assume $M\ge |a|$ and then the Kerr spacetime admits Killing horizons at $r=r_{\rm h(\pm)}$ given by 
\begin{align}
r_{\rm h(\pm)}:=M\pm\sqrt{M^2-a^2}.
\end{align} 
The radius $r=r_{\rm h(+)}$ and $r=r_{\rm h(-)}$ correspond to an event horizon and an inner Cauchy horizon of the Kerr black hole. 
For $M=|a|$, the metric describes an extreme Kerr black hole with a single degenerate Killing horizon at $r=r_{\rm ex}(:=M)$.
As studied in the non-rotating case ($a=0$) in Ref.~\cite{Ovalle:2024wtv}, the metric (\ref{BL}) or (\ref{Doran}) describes a fake Kerr black hole if the mass function $m(r)$ satisfies the following three conditions:
\begin{enumerate}

\item $m(r)=M(\ge |a|)$ holds in the exterior region $r>r_{\rm h(+)}$.

\item $m(r)$ is continuous and $m'(r)$ converges to zero in the limit $r\to r_{\rm h(+)}$ from below.

\item $m''(r)$ is finite in the limit $r\to r_{\rm h(+)}$ from below.
\end{enumerate}
The last two conditions ensure that the metric is at least $C^{1,1}$ on the horizon so that there is no lightlike massive thin shell.

Now we show that at least the DEC is violated inside the event horizon $r<r_{\rm h(+)}$ under the above three conditions.
For the energy-momentum tensor (\ref{matter-rot}) of the Hawking-Ellis type I, equivalent representations of the standard energy conditions are given by~\cite{Maeda:2018hqu,Maeda:2021jdc} 
\begin{align}
\mbox{NEC}:&~~\rho+p_2\ge 0,\label{NEC2}\\
\mbox{WEC}:&~~\rho\ge 0\mbox{~in addition to NEC},\label{WEC2}\\
\mbox{DEC}:&~~\rho-p_2\ge 0\mbox{~in addition to WEC},\label{DEC2}\\
\mbox{SEC}:&~~p_2\ge 0\mbox{~in addition to NEC}.\label{SEC2}
\end{align}
Equation~(\ref{matter-rot}) gives
\begin{align}
\label{EC-set}
\begin{aligned}
&\rho+p_2=\frac{(2m'-rm'')r^2-(2m'+rm'')a^2\cos^2\theta}{\kappa\Sigma^2}=\frac{2m'(r^2-a^2\cos^2\theta)-r\Sigma m''}{\kappa\Sigma^2},\\
&\rho=\frac{2r^2m'}{\kappa\Sigma^2},\qquad \rho-p_2=\frac{2m'+rm''}{\kappa\Sigma},\\
&p_2=-\frac{r^3m''+(2m'+rm'')a^2\cos^2\theta}{\kappa\Sigma^2}=-\frac{r\Sigma m''+2m'a^2\cos^2\theta}{\kappa\Sigma^2}.
\end{aligned}
\end{align}
By $\lim_{r\to r_{\rm h(+)}-0}m'(r)=0$, the NEC requires $m''\le 0$ near the horizon and then the SEC is satisfied.
Then, although the WEC is satisfied if $m'\ge 0$ holds in addition, the DEC is inevitably violated as $\rho-p_2$ becomes negative near the horizon.
We note that the above argument works even when $r_{\rm h(+)}$ is replaced by $r_{\rm h(-)}$ or $r_{\rm ex}$.

\subsection{Three models with different mass functions}

As we have shown that the DEC is inevitably violated near the horizon $r=r_{\rm h(+)}$, we seek a model of fake Schwarzschild-Kerr black holes that satisfies the WEC and SEC everywhere.
We consider the following three different mass functions in the region $r\le r_{\rm h(+)}$;
\begin{align}
m(r)=& M-m_{2+q}\biggl(\frac{r_{\rm h(+)}-r}{r_{\rm h(+)}}\biggl)^{2+q},\label{m-a=0}\\
m(r)=& M-m_{2+q}\frac{(r_{\rm h(+)}-r)^{2+q}}{r_{\rm h(+)}^{1+q}r},\label{m-WEC2}\\
m(r)=& M-m_{2+q}\biggl(\frac{r_{\rm h(+)}-r}{r}\biggl)^{2+q},\label{m-WEC}
\end{align} 
where $M$ and $m_{2+q}$ are positive constants and $q$ is a non-negative constant.
For any of these mass functions, the metric (\ref{Doran}) in the Doran coordinates is $C^{1+q,1}$ at $r=r_{\rm h(+)}$ for integer $q$ and $C^{[2+q]}$ for non-integer $q$.
While we assume $q\ge 0$, the mass functions (\ref{m-WEC2}) and (\ref{m-WEC}) with $q=-1$ become
\begin{align}
m(r)= {\bar M}-\frac{Q^2}{2r},\label{m-KN}
\end{align}
where ${\bar M}:=M+m_{2+q}$ and $Q^2:=2m_{2+q}r_{\rm h(+)}$.
The mass function (\ref{m-KN}) corresponds to the Kerr-Newman solution in the Einstein-Maxwell system.

Hereafter we assume $a^2\ne M^2$ so that $r_{\rm h(-)}<r_{\rm h(+)}$ for simplicity.
Locations of the Killing horizons in the G\"urses-G\"ursey spacetime (\ref{BL}) are determined by $\Delta(r)=0$, which can be written as
\begin{align}
\label{horizon-det}
\begin{aligned}
&(x-1)(x-x_-)=h(x),\\
&h(x):=
\begin{cases}
-2{\bar m}_{2+q}x(1-x)^{2+q} & {\rm for~Eq}.~(\ref{m-a=0})\\
-2{\bar m}_{2+q}(1-x)^{2+q} & {\rm for~Eq}.~(\ref{m-WEC2})\\
-2{\bar m}_{2+q}(1-x)^{2+q}/x^{1+q} & {\rm for~Eq}.~(\ref{m-WEC})\\
\end{cases}
\end{aligned} 
\end{align} 
where $x:=r/r_{\rm h(+)}$, $x_-:=r_{\rm h(-)}/r_{\rm h(+)}$ satisfying $0<x_-<1$, and ${\bar m}_{2+q}:=m_{2+q}/r_{\rm h(+)}$.
The values of $x$ at the intersections of $y=(x-1)(x-x_-)$ and $y=h(x)$ in the domain $x\le 1$ give the locations of the Killing horizons.

\subsubsection{Non-rotating case}

In the non-rotating case ($a=0$), the properties of the fake Schwarzschild black hole for any of the mass functions (\ref{m-a=0})--(\ref{m-WEC}) are similar.
A curvature singularity is located at $r=0$ and Eq.~(\ref{horizon-det}) shows that there is only one inner horizon in the domain $r\in(0,r_{\rm h(+)})$, where $r_{\rm h(+)}=2M$.
In addition, the WEC and SEC are satisfied everywhere inside the fake Schwarzschild black hole as shown below.
Such a Hawking-Ellis type-I matter field (\ref{matter-rot}) with $a=0$ is more physically reasonable than a tachyonic perfect fluid (\ref{T-spacelike}), or equivalently an anisotropic fluid (\ref{Tab-a}), studied in Sec.~\ref{sec:main} since the latter fluid violates the WEC everywhere inside the event horizon.

To check the energy conditions, we compute
\begin{align}
&m'= \frac{(2+q)m_{2+q}}{r_{\rm h(+)}}\biggl(\frac{r_{\rm h(+)}-r}{r_{\rm h(+)}}\biggl)^{1+q},\label{dm1}\\
&m''= -\frac{(1+q)(2+q)m_{2+q}}{r_{\rm h(+)}^2}\biggl(\frac{r_{\rm h(+)}-r}{r_{\rm h(+)}}\biggl)^{q}
\end{align} 
for the mass function~(\ref{m-a=0}),
\begin{align}
&m'=m_{2+q}\frac{(r_{\rm h(+)}-r)^{1+q}[(1+q)r+r_{\rm h(+)}]}{r_{\rm h(+)}^{1+q}r^2},\label{dm2}\\
&m''=-m_{2+q}\frac{(r_{\rm h(+)}-r)^{q}[2r_{\rm h(+)}^2+qr\{(1+q)r+2r_{\rm h(+)}\}]}{r_{\rm h(+)}^{1+q}r^3}\label{ddm2}
\end{align}
for the mass function~(\ref{m-WEC2}), and 
\begin{align}
&m'=\frac{(2+q)m_{2+q}r_{\rm h(+)}(r_{\rm h(+)}-r)^{1+q}}{r^{3+q}},\label{dm3}\\
&m''=-\frac{(2+q)m_{2+q}r_{\rm h(+)}(r_{\rm h(+)}-r)^{q}[(q+1)r_{\rm h(+)}+2(r_{\rm h(+)}-r)]}{r^{4+q}}\label{ddm3}
\end{align} 
for the mass function (\ref{m-WEC}).
With $a=0$, Eq.~(\ref{EC-set}) reduces to
\begin{align}
\begin{aligned}
&\rho+p_2=\frac{2m'-rm''}{\kappa r^2},\qquad \rho=\frac{m'}{\kappa r^2},\\
&\rho-p_2=\frac{2m'+rm''}{\kappa r^2},\qquad p_2=-\frac{m''}{\kappa r}.
\end{aligned}
\end{align}
Equations~(\ref{dm1})--(\ref{ddm3}) show $m'>0$ and $m''<0$ in the domain $r\in(0,r_{\rm h(+)}]$ and therefore the WEC and SEC are satisfied everywhere for any of the mass functions (\ref{m-a=0})--(\ref{m-WEC}).

\subsubsection{Rotating case}

Unlike the non-rotating case, the WEC and SEC are satisfied everywhere inside the fake Kerr black hole only for the mass function (\ref{m-WEC}).
As shown below, the domain of $r$ in the rotating case ($a\ne 0$) is given by $r\in(-\infty,r_{\rm h(+)}]$ for the mass functions (\ref{m-a=0}) and (\ref{m-WEC2}) and by $r\in(0,r_{\rm h(+)}]$ for the mass function (\ref{m-WEC}).
Then, Eq.~(\ref{horizon-det}) shows that there exists only one inner horizon in the region $0<r<r_{\rm h(+)}$ for the mass functions (\ref{m-a=0}) and (\ref{m-WEC}).
For the mass function (\ref{m-WEC2}), there are two inner horizons; one in the region of $0<r<r_{\rm h(+)}$ and the other in the region of $r<0$.

Near $r=0$, the mass functions (\ref{m-a=0})--(\ref{m-WEC}) are expanded as
\begin{align}
&m(r)\simeq (M-m_{2+q})+\frac{(2+q)m_{2+q}}{r_{\rm h(+)}}r+O(r^2),\label{m-a=0-r=0}\\
&m(r)\simeq [M+(2+q)m_{2+q}]-m_{2+q}\frac{r_{\rm h(+)}}{r}-\frac{(1+q)(2+q)m_{2+q}r}{2r_{\rm h(+)}}+O(r^2),\label{m-WEC2-r=0}\\
&m(r)\simeq M-m_{2+q}\frac{r_{\rm h(+)}^{2+q}}{r^{2+q}}\biggl[1-(2+q)\frac{r}{r_{\rm h(+)}}+O(r^2)\biggl],\label{m-WEC-r=0}
\end{align} 
respectively.
In particular, the first two leading terms in the expansion (\ref{m-WEC2-r=0}) are the same form as the mass function (\ref{m-KN}) for the Kerr-Newman solution.
As a result, there is a ring-like curvature singularity at $(r,\theta)=(0,\pi/2)$ for the mass functions (\ref{m-a=0}) and (\ref{m-WEC2}) and the spacetime can be extended into the region of $r<0$ beyond the disk given by $r=0$ with $\theta\ne \pi/2$ as for the Kerr-Newman solution.
On the other hand, $r=0$ is a curvature singularity with any value of $\theta$ for the mass function (\ref{m-WEC}) as the Ricci scalar blows up.

Now we show that the mass function (\ref{m-WEC}) satisfies the WEC and SEC everywhere in the domain $r\in(0,r_{\rm h(+)}]$, while the mass functions (\ref{m-a=0}) and (\ref{m-WEC2}) violate the WEC somewhere in the domain $r\in(-\infty,r_{\rm h(+)}]$.
First, the mass function (\ref{m-a=0}) violates the NEC (and hence the WEC and SEC as well) near $r=0$ with $\theta\ne \pi/2$ since it gives
\begin{align}
&\lim_{r\to 0}(\rho+p_2)=-\frac{2(2+q)m_{2+q}}{\kappa r_{\rm h(+)}a^2\cos^2\theta}<0.
\end{align}
Next, Eq.~(\ref{EC-set}) and Eq.~(\ref{dm2}) for the mass function~(\ref{m-WEC2}) show that $\rho<0$ holds and hence the WEC is violated in the region $r<-r_{\rm h(+)}/(1+q)$.
In contrast, Eqs.~(\ref{dm3}) and (\ref{ddm3}) for the mass function (\ref{m-WEC}) show $m'>0$ and $m''<0$ and therefore $2m'-rm''>0$ is satisfied in the domain $r\in(0,r_{\rm h(+)}]$.
Since they also give
\begin{align}
2m'+rm'' =-\frac{(1+q)(2+q)m_{2+q}r_{\rm h(+)}^2(r_{\rm h(+)}-r)^{q}}{r^{3+q}}<0,
\end{align}
$\rho+p_2\ge 0$, $\rho\ge 0$, and $p_2\ge 0$ are satisfied in the domain $r\in(0,r_{\rm h(+)}]$, while $\rho-p_2<0$ is satisfied in the domain $r\in(0,r_{\rm h(+)})$.
Therefore, everywhere inside the event horizon, the WEC and SEC are satisfied and the DEC is violated.

We lastly comment on the causality violation in those three models.
Among all cases, the fake Kerr black hole described by the G\"urses-G\"ursey metric is free from closed timelike curves only for the mass function (\ref{m-a=0}) with $m_{2+q}=M$. 
In this special case, we have 
\begin{align}
rm(r)= Mr\biggl[1-\biggl(1-\frac{r}{r_{\rm h(+)}}\biggl)^{2+q}\biggl]\ge 0.
\end{align} 
As $rm(r)\ge 0$ is satisfied, the spacetime in the domain $r\in(-\infty,r_{\rm h(+)}]$ can be covered by the Doran coordinates (\ref{Doran}).
It means that a scalar $T:=\eta$ is a time function since its derivative $\nabla_\mu T$ is timelike everywhere, shown as $(\nabla_\mu T)(\nabla^\mu T)=g^{\eta\eta}=-1$.
Then, by Proposition~5 in Ref.~\cite{Maeda:2021jdc}, the spacetime is stably causal and therefore closed timelike curves are absent.

In other cases, closed timelike curves exist inside the fake Kerr black hole.
Consider a closed curve $\gamma$ given by $(t,r,\theta)=(t_0,r_0,\theta_0)$ in the G\"urses-G\"ursey spacetime (\ref{BL}).
The squared norm of its tangent vector $\phi^\mu\partial_\mu=\partial_\phi$ is given by $\phi_\mu \phi^\mu=g_{\phi\phi}$, where
\begin{align}
g_{\phi\phi}=\frac{(r^2+a^2)(r^2+a^2\cos^2\theta)+2a^2m(r)r\sin^2\theta}{\Sigma(r,\theta)}\sin^2\theta.
\end{align} 
By Eqs.~(\ref{m-a=0-r=0})--(\ref{m-WEC-r=0}), it is shown that there is a set of $r_0$ and $\theta_0$ near $r=0$ which gives $g_{\phi\phi}<0$ for any of the mass functions (\ref{m-a=0})-- (\ref{m-WEC}) except for the special case (\ref{m-a=0}) with $m_{2+q}=M$.

For the mass function (\ref{m-a=0}) with $m_{2+q}\ne M$, we obtain 
\begin{align}
g_{\phi\phi}\simeq \left\{a^2+2(M-m_{2+q})r\tan^2\theta\right\}\sin^2\theta
\end{align} 
near $r=0$, which shows that closed timelike curves exist for $m_{2+q}>(<) M$ near $r=0$ in the region of $r>(<)0$ where $\theta$ satisfies
\begin{align}
2(M-m_{2+q})r\tan^2\theta<-a^2.
\end{align} 
For the mass function (\ref{m-WEC2}), we obtain 
\begin{align}
g_{\phi\phi}\simeq \frac{(a^2\cos^2\theta-2m_{2+q}r_{\rm h(+)}\sin^2\theta)+2\{M+(2+q)m_{2+q}\}r\sin^2\theta}{\cos^2\theta}\sin^2\theta
\end{align} 
near $r=0$, which shows that closed timelike curves exist near $r=0$ in the region where $\theta$ satisfies
\begin{align}
\frac{a^2}{2m_{2+q}r_{\rm h(+)}}<\tan^2\theta.
\end{align} 
For the mass function (\ref{m-WEC}), we obtain 
\begin{align}
g_{\phi\phi}\simeq -\frac{2m_{2+q}r_{\rm h(+)}^{2+q}\sin^4\theta}{r^{1+q}\cos^2\theta}<0
\end{align} 
near the singularity $r=0$ with $\theta\ne \pi/2$ and 
\begin{align}
g_{\phi\phi}=-\frac{2a^2m_{2+q}r_{\rm h(+)}^{2+q}}{r^{3+q}}<0
\end{align} 
near the singularity $r=0$ with $\theta= \pi/2$.
Thus, closed timelike curves exist near $r=0$ with any value of $\theta$.

%======================================%
%<<<<<<<<<<<< SECTION I >>>>>>>>>>>>>>%
%======================================%
\section{Concluding remarks}
\label{sec:summary}

In this paper, we have presented several models of {\it fake} Schwarzschild and Kerr black holes.
While the region outside the event horizon of a fake Schwarzschild (Kerr) black hole is described by the Schwarzschild (Kerr) vacuum solution, the region inside the horizon is described by a spherically symmetric (axisymmetric) solution with a matter field.
The metric is required to be at least $C^{1,1}$ on the horizon for regularity, so that there is no lightlike massive thin shell.
In principle, such a fake Schwarzschild (Kerr) black hole cannot be distinguished from the Schwarzschild (Kerr) black hole by observations.

By Proposition~6 in Ref.~\cite{Maeda:2024lbq}, the dynamical region inside a Killing horizon of a static perfect-fluid solution obeying $p=\chi\rho$ for $\chi\in[-1/3,0)$ can be the interior of a fake Schwarzschild black hole.
We have shown in Sec.~\ref{sec:ECs} that the matter field inside such a black hole is not a perfect fluid but an anisotropic fluid and can be interpreted as a spacelike (tachyonic) perfect fluid.
The interior matte field may satisfy the NEC and SEC but violates the WEC at least.
In Ref.~\cite{Maeda:2022lsm}, the author constructed the first exact solution describing a fake Schwarzschild black hole using the Semiz solution for $\chi=-1/5$~\cite{Semiz:2020lxj}.
After a review of this construction in Sec.~\ref{sec:Semiz}, we have presented another model using the Whittaker solution for $\chi=-1/3$ in Sec.~\ref{sec:extension}.

In Sec.~\ref{sec:Semiz}, we have shown that a bifurcation two-sphere of a fake Schwarzschild black hole is regular for $\chi\in[-1/3,0)$, which implies that a counterintuitive configuration shown in Fig.~\ref{Fig-PenroseDiagrams-FakeSemiz2} could be attributed to the violation of the WEC or DEC.
In Sec.~\ref{sec:main2}, we have presented a model of {\it fake Kerr} black holes described by the G\"urses-G\"ursey metric whose interior is filled with a matter field that violates only the DEC and satisfies the WEC and SEC everywhere.
This is achieved at the cost of curvature singularities at $r=0$ with any value of $\theta$.
As explained at the end of Sec.~\ref{sec:Semiz}, these configurations of black holes are not possible under the DEC because they contradict the conservation theorem.

Thermodynamical properties of a fake Schwarzschild (Kerr) black hole are the same as the Schwarzschild (Kerr) black hole.
The entropy of the fake Schwarzschild (Kerr) black hole is identical to that of the Schwarzschild (Kerr) black hole because the areas of their event horizons are the same.
It confirms that the black-hole entropy is independent of the degrees of freedom of the possible configuration inside the horizon.
Also, the temperature of the black hole, which is proportional to the value of the first derivative of a metric function on the event horizon, is the same because the metric is at least $C^{1,1}$ there.
Therefore, the Hawking radiation from a fake Schwarzschild (Kerr) black hole is also the same as the Schwarzschild (Kerr) black hole.
Moreover, the exterior vacuum region of a fake Schwarzschild (Kerr) black hole is dynamically stable as the Schwarzschild (Kerr) black hole~\cite{Regge:1957td,Vishveshwara:1970cc,Zerilli:1970se,Dafermos:2016uzj,Klainerman:2017nrb,Dafermos:2021cbw,Whiting:1988vc} because the boundary conditions for perturbations at the event horizon and spacelike infinity are the same.
In contrast, the stability of the interior region of a fake Schwarzschild or Kerr black hole is a non-trivial problem.

Although fake Schwarzschild and Kerr black holes are rather special configurations, they are allowed under the WEC and SEC.
Our results could suggest that black holes described by a metric that is non-smooth on the horizon are non-negligible.
The problem is worth pursuing whether such black holes are allowed with a physically motivated matter field or in modified gravity.
We leave it for future research.

\subsection*{Acknowledgements}
The author acknowledges gratefully fruitful communications with Jorge Ovalle, who introduced his paper~\cite{Ovalle:2024wtv} and a draft in preparation to the author, which inspired Sec.~\ref{sec:main2} of the present paper.
The author thanks Inyong Cho for pointing out an error in the Penrose diagram of the Whittaker spacetime in the first version of the present paper.
The author also thanks Cristi{\'a}n Mart\'{\i}nez for helpful comments.

\appendix

\section{Energy conditions for a tachyonic perfect fluid}
\label{app:EC-tachyon}

In this appendix, without assuming any spacetime symmetry, we derive equivalent representations of the standard energy conditions for a tachyonic perfect fluid in arbitrary $n(\ge 3)$ dimensions whose energy-momentum tensor is given by Eq.~(\ref{T-spacelike}), namely
\begin{align}
\label{T-spacelike2}
&T_{\mu\nu}=({\tilde\rho}+{\tilde p})s_\mu s_\nu-{\tilde p}g_{\mu\nu},
\end{align}
where $s^\mu$ is a unit spacelike vector satisfying $s_\mu s^\mu=1$~\cite{CordeirodosSantos:2023zja}.
A tachyonic perfect fluid obeying a linear equation of state ${\tilde p}=-{\tilde\rho}$ is equivalent to a cosmological constant $\Lambda=\kappa_n {\tilde p}(=-\kappa_n {\tilde\rho})$.
To show this, projecting the energy-momentum conservation equations $\nabla_\nu T^{\mu\nu}=0$ by $s_\mu$ and $h_{\mu\rho}(=g_{\mu\rho}-s_\mu s_\rho)$ with ${\tilde p}=-{\tilde\rho}$, we respectively obtain
\begin{align}
0=&s_\mu\nabla_\nu T^{\mu\nu}=-s^\nu\nabla_\nu {\tilde p},\\
0=&h_{\mu\rho}\nabla_\nu T^{\mu\nu}=-{h^\nu}_{\rho}\nabla_\nu {\tilde p}.
\end{align}
The above equations show that ${\tilde p}(=-{\tilde\rho})$ is constant everywhere.
%----------------------- lemma ------------------------------%
\begin{Prop}
\label{Pro:EC-tachyonicfluid}
In $n(\ge 3)$ dimensions, the standard energy conditions for a tachyonic perfect fluid~(\ref{T-spacelike2}) are equivalent to 
\begin{align}
\mbox{NEC}:&~~{\tilde\rho}+{\tilde p}\ge 0,\label{NEC-tachyon}\\
\mbox{WEC}:&~~{\tilde p}\ge 0\mbox{~in addition to NEC},\label{WEC-tachyon}\\
\mbox{DEC}:&~~{\tilde\rho}-{\tilde p}\le 0\mbox{~in addition to WEC},\label{DEC-tachyon}\\
\mbox{SEC}:&~~{\tilde\rho}-{\tilde p}\ge 0\mbox{~in addition to NEC}.\label{SEC-tachyon}
\end{align}
\end{Prop}
{\it Proof}. 
We consider orthonormal basis vectors ${E}^\mu_{(a)}=({E}^\mu_{(0)},{E}^\mu_{(1)},\cdots,{E}^\mu_{(n-1)})$ that satisfy
\begin{equation}
{E}^\mu_{(a)}{E}_{(b)\mu}=\eta_{(a)(b)}=\mbox{diag}(-1,1,\cdots,1).
\end{equation}
The metric $g_{\mu\nu}$ in the spacetime is given by $g_{\mu\nu}=\eta_{(a)(b)}E^{(a)}_{\mu}E^{(b)}_{\nu}$.
Since $s^\mu$ is spacelike, we can set ${E}^\mu_{(1)}=s^\mu$ without loss of generality.
Then, orthonormal components of $T_{\mu\nu}$ in the orthonormal frame are computed to give
\begin{align}
T^{(a)(b)}=&\eta^{(a)(c)}\eta^{(b)(d)}T_{\mu\nu}{E}^\mu_{(c)}{E}^\nu_{(d)} \nonumber \\
=&\mbox{diag}({\tilde p},{\tilde\rho},-{\tilde p},\cdots,-{\tilde p}).
\end{align}
This is the following canonical form of the Hawking-Ellis type-I energy-momentum tensor
\begin{equation} 
T^{(a)(b)}=\mbox{diag}(\mu, p_1,p_2, \cdots,p_{n-1})
\end{equation}
with $\mu={\tilde p}$, $p_1={\tilde\rho}$, and $p_i=-{\tilde p}~(i=2,3,\cdots,n-1)$.
Then, by Propositions~1--5 in Ref.~\cite{Maeda:2018hqu}, equivalent representations of the standard energy conditions are given by Eqs.~(\ref{NEC-tachyon})--(\ref{SEC-tachyon}).
\qed
%----------------------- lemma ------------------------------%

\noindent
It is noted that, unlike a timelike perfect fluid, the inequality (\ref{SEC-tachyon}) for the SEC does not depend on the number of spacetime dimensions $n$.

Proposition~\ref{Pro:EC-tachyonicfluid} shows that a tachyonic perfect fluid~(\ref{T-spacelike2}) satisfies all the standard energy conditions if and only if ${\tilde p}={\tilde\rho}(\ge 0)$.
In fact, such a fluid is equivalent to a massless scalar field\footnote{It was shown in Ref.~\cite{Santos:1993fn} that a massive scalar field $\phi$ with spacelike $\nabla_\nu \phi$ is equivalent to a tachyonic fluid.}.
%----------------------- lemma ------------------------------%
\begin{Prop}
\label{Prop:tachyonicfluid-scalar}
A tachyonic perfect fluid~(\ref{T-spacelike2}) obeying ${\tilde p}={\tilde\rho}(>0)$ is equivalent to a massless scalar field $\phi$ whose derivative is spacelike and non-vanishing.
\end{Prop}
{\it Proof}. 
The energy-momentum tensor of a massless scalar field is given by 
\begin{align}
&T_{\mu\nu}=(\nabla_\mu \phi)(\nabla_\nu \phi)-\frac12g_{\mu\nu}(\nabla\phi)^2. \label{T-scalar}
\end{align}
In a region where $(\nabla\phi)^2>0$ holds, Eqs.~(\ref{T-scalar}) and (\ref{T-spacelike2}) are identical, shown by the following identifications:
\begin{align}
{\tilde p}={\tilde\rho}\equiv \frac12(\nabla\phi)^2,\qquad s_\mu\equiv \frac{\nabla_\mu\phi}{\sqrt{(\nabla\phi)^2}}.
\end{align}
\qed
%----------------------- lemma ------------------------------%

\noindent
Equivalent representations of the standard energy conditions for a minimally coupled scalar field with an arbitrary self-interacting potential $V(\phi)$ are available in Corollary~6 in Ref.~\cite{Maeda:2022vld}.

\section{Topological generalization of the Whittaker solution}
\label{app:topo}

In this appendix, we perform the same analysis presented in Sec.~\ref{sec:extension} to the topological generalization of the Whittaker solution with an arbitrary two-dimensional maximally symmetric base manifold in the presence of $\Lambda$, which was given by Hinoue, Houri, Rugina, and Yasui as a non-rotating limit of the Wahlquist solution~\cite{Hinoue:2014zta}. 
The metric is given in the conventional coordinates $(T,r,\theta,\phi)$ as
\begin{align}
\label{whittaker-g}
&\D s^2=-f\D T^2+\frac{\D r^2}{(1-\alpha r^2)f}+r^2\D\Omega_{(k)}^2
\end{align}
with
\begin{align}
\label{whittaker-f+}
&f(r)=k-\frac{2M\sqrt{1-\alpha r^2}}{r} -\frac{\Lambda}{\alpha}\biggl[1-\frac{\mbox{arcsin}(\sqrt{\alpha} r)\sqrt{1-\alpha r^2}}{\sqrt{\alpha} r}\biggl]
\end{align}
for $\alpha>0$ and 
\begin{align}
\label{whittaker-f-}
&f(r)=k-\frac{2M\sqrt{1-\alpha r^2}}{r} -\frac{\Lambda}{\alpha}\biggl[1-\frac{\mbox{arcsinh}(\sqrt{|\alpha|} r)\sqrt{1-\alpha r^2}}{\sqrt{|\alpha|} r}\biggl]
\end{align}
for $\alpha<0$, where $M$, $\alpha$, and $k(=1,0,-1)$ are parameters\footnote{There is a typo in Eq.~(A8) in Ref.~\cite{Hinoue:2014zta}.} .
Here $\D\Omega_{(k)}^2$ is the line element on the two-dimensional maximally symmetric base manifold given by 
\begin{align}
\D\Omega_{(k)}^2 = \left\{
\begin{array}{ll}
\D\theta^2+\sin^2\theta\D\phi^2 & (k=1)\\
\D\theta^2+\theta^2\D\phi^2 & (k=0)\\
\D\theta^2+\sinh^2\theta\D\phi^2 & (k=-1)
\end{array}
\right..
\end{align}
The metric given by Eqs.~(\ref{whittaker-g})--(\ref{whittaker-f-}) and a perfect fluid~(\ref{whittaker-matter}) solve the following Einstein equations
\begin{align}
\label{EFE-lambda}
\begin{aligned}
&G_{\mu\nu}+\Lambda g_{\mu\nu}=\kappa T_{\mu\nu},\\
&T_{\mu\nu}=(\rho+p)u_\mu u_\nu+pg_{\mu\nu}.
\end{aligned}
\end{align}
For $\alpha=0$, the solution (\ref{whittaker-g}) reduces to the topological Schwarzschild-(anti-)de~Sitter solution.

In the comoving quasi-global coordinates $(t,x)$ defined by Eq.~(\ref{trans+}), we have $H(x)$, $u^\mu$, $\rho$, and $p$ given by Eq.~(\ref{matter-x}).
A Killing horizon $r=r_{\rm h}$ determined by $f(r_{\rm h})=0$ corresponds to $x=x_{\rm h}$ determined by $H(x_{\rm h})=0$.
For $\alpha>0$, Eq.~(\ref{trans+}) gives $r(x)$ as Eq.~(\ref{r+}), while the other metric function is given as 
\begin{align}
\label{Hr+app}
&H(x)=\frac{1}{\omega}\biggl\{k-\frac{2\sqrt{\alpha}M}{\tan(\sqrt{\alpha\omega}x)} -\frac{\Lambda}{\alpha}\biggl(1-\frac{\sqrt{\alpha\omega}x}{\tan(\sqrt{\alpha\omega}x)}\biggl)\biggl\}.
\end{align} 
For $-1/(4M^2)<\alpha<0$, Eq.~(\ref{trans+}) gives $r(x)$ as Eq.~(\ref{r-}), while the other metric function is given as 
\begin{align}
\label{Hr-app}
H(x)=&\frac{1}{\omega}\biggl\{k-\frac{2\sqrt{|\alpha|}M}{\tanh(\sqrt{|\alpha|\omega}x)} -\frac{\Lambda}{\alpha}\biggl(1-\frac{\sqrt{|\alpha|\omega}x}{\tanh(\sqrt{|\alpha|\omega}x)}\biggl)\biggl\}.
\end{align} 
With non-zero $\Lambda$, algebraic equations $f(r_{\rm h})=0$ and $H(x_{\rm h})=0$ cannot be solved explicitly.

Now we choose $\omega$ such that $r'(x_{\rm h})=1$, namely
\begin{align}
\omega=\frac{1}{1-\alpha r_{\rm h}^2}.
\end{align} 
Then, using 
\begin{align}
M=&\frac{r_{\rm h}}{2\sqrt{1-\alpha r_{\rm h}^2}} \biggl\{k-\frac{\Lambda}{\alpha}\biggl[1-\frac{\mbox{arcsin}(\sqrt{\alpha} r_{\rm h})\sqrt{1-\alpha r_{\rm h}^2}}{\sqrt{\alpha} r_{\rm h}}\biggl]\biggl\}
\end{align}
to remove $M$, we obtain
\begin{align}
\label{ddg-h-app}
\begin{aligned}
&H'(x_{\rm h})=\frac{k-\Lambda r_{\rm h}^2}{r_{\rm h}},\qquad H''(x_{\rm h})=-\frac{2k}{r_{\rm h}^2},\\
&r'(x_{\rm h})=1,\qquad r''(x_{\rm h})=-\frac{\alpha r_{\rm h}}{1-\alpha r_{\rm h}^2}.
\end{aligned} 
\end{align} 
By the same argument as that in Sec.~\ref{sec:extension}, two generalized Whittaker spacetimes $({\cal M}_\pm^4,g^\pm_{\mu\nu})$ with the parameters $(\alpha,M,k,\Lambda)=(\alpha_\pm,M_\pm,k,\Lambda)$ can be attached regularly on a Killing horizon without a lightlike thin shell if the first junction condition $r_{\rm h}(\alpha_+,M_+)=r_{\rm h}(\alpha_-,M_-)$ is satisfied.
Although this condition cannot be written down explicitly, regularity of the Killing horizon is ensured by Eq.~(\ref{ddg-h-app}), which shows that the metric is at least $C^{1,1}$ on the horizon.

%======================================%
%<<<<<<<<<<<<< REFERENCES >>>>>>>>>>>>>%
%======================================%

\end{document}